\documentclass{article}
\usepackage{amssymb}
\usepackage{latexsym}
\begin{document}

\title{Harrison Cohomology and \\Abelian Deformation Quantization\\
on Algebraic Varieties}

\author{Christian Fr\o nsdal\thanks{e-mail:\,\texttt{fronsdal@physics.ucla.edu};
\quad Date: July 10, 2001} \\
{Department of Physics \& Astronomy, University of California,}\\
{Los Angeles, CA 90095-1547, USA }}
\markboth{Christian Fr\o nsdal}{
Abelian Deformation Quantization on Algebraic Varieties}
\date{}
\maketitle

\begin{abstract}
 Abelian deformations of ordinary algebras of
functions are studied. The role of Harrison cohomology in classifying such
deformations is illustrated in the context of simple examples chosen for
their relevance to physics.  It is well known that Harrison cohomology is
trivial on smooth manifolds and that, consequently,  abelian
$*$-products on such manifolds are trivial to first order in the
deformation parameter. The subject is nevertheless interesting;  first
because varieties with singularities appear in the physical context and
secondly, because deformations that are trivial to first order are not
always (indeed not usually) trivial as exact deformations. We investigate
cones, to illustrate the situation on algebraic varieties, and we point
out that the coordinate algebra on  (anti-) de Sitter space is a nontrivial
deformation of the coordinate algebra on Minkowski space  -- although
both spaces are smooth manifolds. 
\end{abstract}


\section{Introduction}  
Deformation quantization has become a dynamic subject in mathematics
as well as in physics. The general setting is a (commutative and
associative) algebra of functions, and the most important case is the
algebra of differentiable functions on a symplectic space, the observables
of classical mechanics. The problem is to construct an associative algebra,
generally non commutative, on the same space, as a formal or exact
deformation of this algebra. 

The Poisson bracket of classical mechanics plays an important role.
A deformation `in the direction of the Poisson bracket' is a formal
associative product on the algebra of formal power series,
$$
f*g = fg + \sum_{n\geq 0}\hbar^nC_n(f,g),
$$
such that $C_1(f,g) = \frac{1}{2}\{f,g\}$. The ordinary product $fg$ and the 
Poisson bracket $\{f,g\}$ are extended to formal power series in $\hbar$.
The emphasis on this type deformation can be explained, not only by the
physical applications, but by the fact that, on a smooth manifold,
every associative deformation is equivalent to one with $C_1$
antisymmetric. If $C_1$ is linear in $df$ and in $dg$,  and
antisymmetric,  then by the associativity of the $*$-product it follows
that $f,g \mapsto C_1(f,g)$ is a Poisson bracket. 

The existence of a Poisson bracket turns the manifold into a Poisson
manifold but not, in general, into a symplectic space. The existence and
the construction of $*$-products on an arbitrary Poisson manifold has
recently been investigated, with great success \cite{K,T}.

Abelian deformations represent another type of generalization. Such
deformations are not trivial. In the first place, though it is true that on a
smooth manifold every first order, abelian deformation is trivial, the
same is not true of exact deformations. In the second place, it is
interesting to go beyond the context of smooth manifolds; and 
in particular, to attempt the construction of
$*$-products on algebraic varieties with singularities.

Abelian $*$-products have turned up in at least one physical context:
in attempts to quantize Nambu mechanics\cite{DFST}, where methods 
inspired by second quantization had to be used. First order abelian 
deformations are classified by Harrison cohomology\cite{B,H}, 
which is trivial on smooth manifolds.
In order to overcome this difficulty attempts have been made \cite{N,P} to
generalize algebraically the usual notion \cite{GS} of (Gerstenhaber) 
deformations, taking a deformation parameter which acts by automorphisms
(to the right and to the left) on the original algebra.

Dealing with algebras of functions (in particular, polynomials)
on manifolds or varieties, 
it is natural to ask what types of singularities would be required for
the existence of a non trivial, commutative $*$-product.  
This paper was planned as a study of simple examples of 
algebraic varieties, notably cones.
However, in the course of this study, it became clear that interesting, 
non trivial deformations exist even on smooth manifolds (including
$\mathbb{R}^n$). These are exact deformations: the deformed product is exact
to all orders of $\lambda$, but the real surprise was the realization that
there are exact, abelian, associative $*$-products, on smooth manifolds, 
of the form
$$
f*g = fg + \lambda C(f,g)
$$ 
(with no terms of higher order in $\lambda$),  that are non trivial
as exact deformations.

\smallskip

Physical applications, some that are immediate and some that are more
remote, include the following:
\begin{itemize}
\item[(a)] The conformal anomaly, briefly discussed in Section 2.6.
\vspace*{-2.2mm}

\item[(b)] A cohomological classification of operator product expansions
in quantum field theories.
\vspace*{-2.2mm}

\item[(c)] A view of classical (anti-) de Sitter field theory as an abelian
deformation of Min\-kowski field theory; see Section 3.  We hope that this
may lead to a better understanding of some difficult aspects of anti-de
Sitter physics.
\vspace*{-2.2mm}

\item[(d)] Within the program of geometric  or $*$-product quantization on
co-adjoint orbits, there are unsolved difficulties in the case of conic
orbits. Most interesting is the quantization of Keplerian systems.
\vspace*{-2.2mm}

\item[(e)] One may look into the possibility of finding central extensions of
Virasoro algebras on algebraic varieties,  for example, on cones.
\end{itemize}

Section 2 initiates our study by detailed calculations on very simple
examples.  This paper considers only algebraic varieties, $\mathbb{R}^n
$or subvarieties of $\mathbb{R}^n$ defined by the vanishing of a finite set
of polynomials. For physical applications it may be that the most
interesting algebraic varieties are the cones,
$$
\mathbb{R}^n/g,\quad g = \sum g^{ij}x_ix_j,\quad g^{ij} \in \mathbb{R},
\quad i,j =1,...n. 
$$
The simplest case, $n = 2$ and $g = x^2-y^2 = uv$ is investigated in
Section 2.3. 

Section 3 deals in great detail with a deformed, abelian algebra of
functions on Min\-kowski space that is isomorphic to an algebra of
functions on (anti-) de Sitter space.

The remaining sections study Hochschild (including Harrison) homology 
and cohomology,  on cones of any dimension. The main purpose is to develop an
understanding of the relationship between singularities and cohomology
and, in particular, the role of singularities within a program of
deformation quantization.

\section{First Examples}  
 
  We shall investigate the Harrison  cohomology for a commutative
algebra, in the simplest context.

\smallskip

\noindent\textit{2.1. Harrison cohomology  and abelian $*$-products.}\\
Harrison's complex  for a commutative and associative algebra 
${\mathcal{A}}$ is a subcomplex of the Hochschild complex. The cochains are 
valued in the unital augmentation of the algebra itself and familiar formulas
for the differential apply. Thus if $E$ is a one-form, then
$$
dE(f,g) = fE(g) -E(fg) + E(f)g,
$$
which vanishes for derivations, and if $C$ is a two-form, then
$$
dC(f,g,h) = fC(g,h) - C(fg,h) + C(f,gh) - C(f,g)h.
$$
A first order, abelian deformation of  (the product of) the algebra  is a
new product on the same space, a $*$-product,
$$
f*g = fg + \lambda \,C(f,g),\quad C(f,g) = C(g,f),\quad f,g \in {\mathcal{A}},
$$
such that, to first order in $\lambda$, $ (f*g)* h = f*(g*h)$. Here
$\lambda$ is a formal parameter and it is implicit that the original
algebra must be extended to the commutative algebra of polynomials in
$\lambda$. 
 The condition of associativity  is equivalent, to first order in
$\lambda$, to the condition that $C$ be closed, $dC = 0$.

We continue to discuss first order deformations, until further notice.
If there is a one-form $E$, such that $C = dE$, then the deformation is
said to be trivial. The reason for this is as follows. For 
$f \in {\mathcal{A}}$, let $f_\lambda := f + \lambda E(f)$. 
Then to first order in $\lambda$,
$$
 f_\lambda g_\lambda = (fg)_\lambda + \lambda C(f_\lambda,g_\lambda),
$$
so that the first order mapping  $f \mapsto f + \lambda E(f)$ 
is a first order isomorphism from ${\mathcal{A}}$  to the deformed algebra. 

\smallskip

\noindent\textit{2.2. A trivial $*$-product.} \\
Let ${\mathcal{A}}$ be the coordinate algebra of $\mathbb{R}^n$,
$$
A =\mathbb{C}[x_1,...,x_n],
$$ 
 and
${\mathcal{A}}_\lambda$ the same with a deformed product.  Notice that
this algebra does not have a unit. To prove that every deformation
 is trivial to all orders  
(we are here talking about formal deformations by infinite series that 
may or not converge for any $\lambda \in \mathbb{C}$), 
let  
$$
\Phi: {\mathcal{A}} \rightarrow {\mathcal{A}},\quad \Phi(f*) = f,
$$
where $f*$ is obtained from $f$ by replacing $x_ix_j\ldots$ by 
$x_i*x_j*\ldots\,\,$. This map takes a Poincar\'e-Birkhoff-Witt basis for
${\mathcal{A}_\lambda}$ to   a Poincar\'e-Birkhoff-Witt basis for
${\mathcal{A}}$ and is an algebra isomorphism.  To relate this to the
foregoing discussion of first order deformations, consider the one-form 
$$
\lambda E(f) = (f*)-f,\quad f* = \Phi^{-1}(f).
$$
  We have $((fg)*) = (f*)*(g*)$ and  to first order in $\lambda$,
\begin{eqnarray*}
\lambda dE(f,g) &=& [(f*)*(g*) - fg] - f[(g*)-g] - [(f*)-f]g \\
&=& f*g-fg \, = \, \lambda C(f,g).
\end{eqnarray*}

For a more concrete example take $n = 1, C(f,g) = f_2g_2$, where 
$f = f_1 + xf_2$ is the unique decomposition of $f$ in terms of two even
polynomials $f_1,f_2$. In this case $C = dE$ with
$$
E(f) = {-\frac{1}{2x}}(\partial f_1 + x\partial f_2) = 
{-\frac{1}{2x}}\partial f + {\frac{1}{2x}}f_2.
$$ 
This is  a polynomial,  though each term separately is not.   But the
first term is formally closed (a derivation), and
$$
dE(f,g)  = {-\frac{1}{2x}}\bigl(g_2f_1 + g_2f_1 - fg_2 - gf_2\bigr)
= f_2g_2  = C(f,g).
$$ 

To find a nontrivial deformation we must go beyond 
this example, and one way to go is to introduce relations.

\smallskip

\noindent\textit{2.3. Algebras defined by relations.}\\
Let $M$ be the algebraic variety
$$
M = \mathbb{R}^n/R,
$$
where $R$ is a set of polynomial relations. Let ${\mathcal{A}}$
be the coordinate algebra of $M$, namely
$$
{\mathcal{A}} = \mathbb{C}[x_1,...,x_n]/R.
$$ 
Let ${\mathcal{A}}_\lambda$ be a deformed algebra and let 
$\Phi: {\mathcal{A}}\rightarrow {\mathcal{A}}$ be the map introduced above,  
that takes $f*\mapsto f$. Let $R_* $ be the relations of 
${\mathcal{A}}_\lambda$ and $R_\lambda = \Phi R_*$; then the two algebras 
are isomorphic if there is an invertible mapping  on $\mathbb{R}^n$ that 
takes $R_\lambda$ to $R$.

\smallskip

\noindent\textit{2.4. First non trivial example.} \\
Let $n=1$, $N$ a natural number, $r\in \mathbb{R}$, and $R =  x^N-r$.  
The Poincar\'e-Birkhoff-Witt basis of ${\mathcal{A}} = \mathbb{C} [x]/R$ is 
$\{1,x,x^2,\ldots, x^{N-1}\}$. Define a deformed product by setting
$$
x^k*x^l = \biggl\{{\hskip-0.1cm x^{k+l}  \hskip17mm , 
\,\,\,\,\,\,k+l <N,\atop
x^{k+l}+\lambda x^{k+l-N} \,\,\, , \,\,\,\,\,\, k+l \geq N.}
$$
Then $R_* = (x*)^N -(r+ \lambda), \,\,R_\lambda = x^N-(r+\lambda)$
and a trivializing map is given by 
$$
x \mapsto (1+\lambda/r)^{1/N}x.
$$
Two things can go wrong. (1) This map does not exist if $r = 0$, and in
that case the deformation is not trivial.  (2)  If $r \neq 0$, then the
trivializing map is an infinite power series in $\lambda$.  
 
We have $f*g = fg + \lambda C(f,g)$, with
 $$
C(x^k,x^l) =  \biggl\{ {0 \hskip8.5mm,\hskip4mm k+l < N,\atop
  x^{k+l-N} , \quad k+l \geq N.}
$$
There is no need to add higher order corrections in $f*g$; the product
as it stands is associative,  to all orders in $\lambda$.  We have $C = dE$,
with $E(x^k) = (k/Nr)x^k,\quad k = 1,..., N-1$.
\vspace*{1mm}

\noindent\textit{Conclusion}: Although $C$ is  first order, exact,  
and the $*$-product is associative to all orders, nevertheless an
infinite power series was needed to trivialize it.  
In this particular case the appearance of infinite series is more or 
less innocuous, but that is not always the case, as we shall see.  
 
\smallskip

\noindent\textit{2.5. Conic sections.} \\
Let ${\mathcal{A}} = \mathbb{C}[x,y]/R$, with $R = y^2-x^2 - r^2$. 
There is a unique decomposition 
$$f = f_1 + yf_2,$$
where $f_1,f_2$ are polynomials in $x$. Take $C(f,g) = f_2g_2$. This 
two-form is closed. The deformed relation is $ y^2 -x^2 -( r^2 + \lambda)$. 
These relations define equivalent algebraic varieties if $r^2 \neq 0$
(when $r\neq 0$ there is a neighbourhood of $\lambda =0$ 
in which the two varieties are equivalent), 
so in this case the deformation is trivial. But if $r = 0$, when the 
initial variety is a cone and has a singularity at
$ x = y = 0$, the deformation is not trivial. 
We shall return to cones in Section 4.

The polynomial one form
$$
E(f) = \frac{1}{2r^2}[x\partial f_1 + yf_2 + xy\partial f_2]  =
\frac{x}{2r^2}\partial f + \frac{1}{2y}f_2 
$$ 
(in the last expression $\partial y = x/y$) is formally cohomologous to
$E'(f) := \frac{1}{2y}f_2$, and
$$
dE(f,g) = dE'(f,g) = -\frac{1}{2y}[f_1g_2 + f_2g_1 - fg_2 - f_2g]
= f_2g_2.
$$  
This   shows that the first order deformation is trivial when $r^2 \neq 0$. 
The derivation included in the definition of $E$, to eliminate the
non polynomiality of $E'$, does not exist when $r^2 = 0$.  We notice
that the obstruction is the constant term in $f_2$; in other words the
value of $C(f,g)$ at the origin.
 
\smallskip

\noindent\textit{2.6. Application, conformal anomaly ?} \\
We may consider the algebra of polynomials on Dirac's cone.  Of interest is 
the projective cone, one studies functions with a fixed degree of 
homogeneity, normally a negative integer. In scalar field theory the degree
is -1.  The projective cone can be covered by two charts. If $y$ is a set 
of coordinates for $\mathbb{R}^6$ then a pair of  charts is given by 
$x^\pm_\mu = y_\mu/(y_5\pm y_6), ~\mu = 1,2,3,4$,
each mapping most of the projective cone onto Minkowski space. To define a
function on compactified Minkowski space we may take a pair of functions on
Minkowski space, with the relation $f_1(x) = f_2({x}/{x^2})$.
The ordinary product would be $fg= (f_1,f_2)(g_1,g_2) = (f_1g_1,f_2g_2)$.
It can probably be deformed, and such deformations may perhaps relate to the
conformal anomaly, but that will be left for another time.

\section{Field theory on  anti-de Sitter space}  

Let ${\mathcal{A}} = \mathbb{C}[x^0,... \,,x^3]$, the coordinate algebra of 
polynomial functions on 
Minkowski space, without constant term. The metric 
is $g_{ij}dx^idx^j = (dx^0)^2 - \sum_{i = 1}^3(dx^i)^2$.
For every $a\in {\mathcal{A}}$ let $ a = a_+ + a_-$ be the decomposition into
an even and an odd part. We regard the algebra as the space of pairs,
$$
A = \{(a_+,a_-)\},
$$ with the product
$$
AB = (a_+b_+ + a_-b_-, a_+b_- + a_-b_+),
$$
or equivalently as the even subalgebra of  
$\mathbb{C}[x^0, \ldots,x^3,y]/(y^2-1)$, with elements
$$
a = a_+ + ya_-.
$$

We deform the product, setting
$$
A*B = AB - \rho x^2a_-b_-,\quad x^2 := g_{ij}x^ix^j,\quad \rho > 0.
$$
The deformed algebra is the even subalgebra of
$$
\mathbb{C}[x^0,\ldots,x^3,y]/(y^2-1+\rho x^2),
$$ which is a coordinate algebra  of polynomial functions on 
anti-de Sitter space.

Referring back to  Section 2.3, we notice a great deal of similarity. The
above $*$-product is associative to all orders in $\rho$, no higher order
terms are needed. Furthermore, $C(a,b) = \rho x^2a_-b_-$  defines an
exact two-form, $C = dE$ with $E(a) = (\rho/2)x^2a_-$. But the
``trivializing" map takes $a \mapsto a_+ + y\sqrt{1+\rho x^2}a_-$, and
here we find an infinite power series in the generators of the algebra, not
just in the parameter. And this series is not entire but has singularities,
for any $\rho$ different from zero. There is no sensible point of view 
that would allow us to regard the two algebras as equivalent.
\bigskip

\noindent\textit{Conclusion.}  The algebra of coordinate functions on 
anti-de Sitter space is a nontrivial deformation of that of Minkowski space; 
this in spite of the fact that both spaces are smooth manifolds. 
The deformations are not classified by Harrison cohomology, trivial in 
this case. It is true that Harrison cohomology does classify first order 
deformations, but it must be understood that, even if 
$a*b = ab + \lambda C(a,b)$, with no higher order terms, and this is 
associative to all orders, trivialization is not localized at finite order. 
Vanishing of $\mathrm{Harr}^2$ merely tells us that the term of first order 
in $\lambda$ can be removed, to be replaced by terms of higher order.

Abelian $*$-products thus appear in two different ways. In the cases
when $\mathrm{Harr}^2 = 0$ the obstructions are singularities of the
``trivializing" map. But this does not diminish our interest in trying to
understand those abelian $*$-products that owe their non-triviality to
the existence of Harrison cohomology, and that is the subject of the rest
of this paper.

\section{Hochshild homology of the cone  uv = 0} 

This section and the next take up the study of the cone algebra introduced
in 2.5, with a more convenient choice of coordinates. 

Let $M = \{u,v \in \mathbb{R}^2, uv = 0\}$, and let 
${\mathcal{A}}$ be the commutative algebra
$$
{\mathcal{A}} = \mathbb{C}[u,v]/uv.
$$
We study the complex 
$$
{{\hskip-3mm\partial\hskip7mm\partial}\atop 
{0\leftarrow C_1 \leftarrow C_2\,\,\cdots}}
$$
where $C_n \in   {\mathcal{A}}^{\otimes n} $, and
$$
\partial(a_1\otimes a_2\otimes \cdots \otimes a_n) = a_1a_2\otimes
a_3\otimes \cdots - a_1\otimes a_2a_3\otimes a_4\otimes \cdots +
\cdots +
(-1)^n a_1\otimes \cdots \otimes a_{n-1}a_n.
$$
Let $Z_n,B_n$ denote the subspaces of closed, respectively exact
subspaces of $C_n$:
$$
Z_n = \{ a\in C_n, \partial a = 0\},\quad B_n = \partial C_{n+1}.
$$

\noindent\textbf{Theorem 4.1} \textit{
The  Hochschild homology $H_n = Z_n/B_n$
of this complex is of dimension 2 for $n \geq 1$, and is generated as a
vector space by $u\otimes v\otimes u\otimes \cdots$ and $v\otimes
u\otimes v\otimes \cdots\,$.}

\noindent\textit{Proof.} For every $n$-chain $a$ we have
$$
\partial(u\otimes a) = ua_1\otimes a_2 \otimes \cdots - u\otimes da,
$$
therefore every $n$-chain with initial factor $ua_1$ is cohomologous
with one with initial factor $u$, and every $n$-chain is cohomologous
with $u\otimes a + v\otimes b$. Now suppose that $u\otimes a$ is closed. Then
$$
u\otimes a = u\otimes vb_1\otimes b_2 \otimes \cdots \approx  u 
\otimes v\otimes db,
$$
and so on. Since no $n$-chain $a$ with each factor $a_i$ linear in
$u,v$ is exact, the theorem is proved. \hfill $\Box$

\section{Hochschild cohomology of the cone uv = 0} 

The cochains are valued in the unital augmentation ${\mathcal{A}_1}$ 
of ${\mathcal{A}}$,  and
$$
dC(a_1,\ldots,a_n)= a_1C(a_2,...,a_n) - C(da) +(-1)^nC(a_1,\ldots,a_{n-1})a_n.
$$

For any $n$-cochain $C$, let $C'$ denote the restriction 
of $C$ to \textsl{closed}, \textsl{linear} chains.
Thus $C'$ is defined by the values $C'(a) = C(a)$, with $a = u\otimes v
\otimes u\otimes \cdots$ and $a = v\otimes u\otimes v\otimes \cdots$.
The set of all restricted cochains form a complex with differential
$$
d'C'(a) = a_1C'(a_2,...a_n) + (-1)^nC'(a_1,...,a_{n-1})a_n
$$
A  restricted $2n$-cochain is closed if
$$ 
C'_{2n}(u,v,...,v) = C'_{2n}(v,u,...,u), 
$$
and exact if it is valued in ${\mathcal{A}}$. A restricted $2n+1$-form is
closed if 
$$
  vC'_{2n+1} (u,v,...,u) + uC'_{2n+1}(v,u,...,v) = 0,
$$
which implies that $C(u,...,u) = u\alpha (u),\, C(v,...,v) = v\beta (v)$,
with $\alpha(u), \beta(v)$    in the unital augmentation of ${\mathcal{A}}$,
and it is exact if  $n > 1$ and $\alpha + \beta \in{\mathcal{A}}$. The
cohomology class of $2n$-cochain is  the value of $C(u,v,u, \cdots)$ at
the origin.

\noindent\textbf{Proposition 5.1.} \textit{
For every closed $n$-cochain $\tilde C$ of the restricted complex there 
is a closed Hochschild cochain $C$ such that the restriction $C'$ of $C$ 
is equal to $\tilde C$.}

\noindent\textit{Proof.}
To find such $C$ we begin by setting $C' = \tilde C$ for
linear, closed arguments, then attempt to define $C$ for nonlinear
arguments by induction in the polynomial degree, using
$$
dC(a)  = a_1C(a_2,...) - C(da) + (-1)^nC(...,a_{n-1})a_n=0,
$$
with the observation that the argument of the middle term is of higher
total polynomial degree  than the others.  The obstruction is $da = 0$ 
with $a$ not linear; thus $a = db$.  For an inductive proof, we must
show that, for $a=db$, the last relation holds by virtue of similar 
relations involving arguments of lower order.  In fact,
\begin{eqnarray*}
dC(db) &=& b_1b_2C(b_3,..., b_{n+1}) - b_1C(db^+)\\
&& +(-1)^n\bigl(C(db^-)b_{n+1}+ (-1)^{n+1}C(b_1,...,b_{n-1})b_nb_{n+1}\bigr),
\end{eqnarray*}
where $b^+ = b_2\otimes \cdots \otimes b_{n+1}$ and 
$b^- = b_1\otimes \cdots b_{n}$. By the induction hypothesis the second
and third term are
\begin{eqnarray*}
&-&b_1\biggl(b_2C(b_3,\cdots,b_{n+1})
+(-1)^nC(b_2,\cdots,b_n)b_{n+1}\biggr)\\
&+&(-1)^n\biggl(b_1C(b_2,\cdots,b_n) +
(-1)^nC(b_1,\cdots,b_n)\biggr)b_{n+1},
\end{eqnarray*}
and substitution gives the desired result, zero. \hfill $\Box$
\medskip

\noindent\textbf{Proposition 5.1.} \textit{
If $C$ is any closed cochain, and its restriction is exact, then $C$ is exact.}

\noindent\textit{Proof.} We must show that there is $E$ such that
$$
C(a) = a_1E(a_2,...) - E(da) + (-1)^nE(...,a_{n-1})a_n.
$$
By hypothesis this is true if $a$ is linear and closed, and that establishes
a basis for induction in the total polynomial degree of $a$. The
obstruction is again $a = db$, and  it is necessary to show that $C(db)$ is
the same as
\begin{eqnarray*}
&&b_1b_2E(b_3,\cdots ,b_{n+1}) - b_1E(db^+)\\
&&+(-1)^n\biggl(E(db^-)b_{n+1} +
(-1)^nE(b_1,\cdots,b_{n-1})b_nb_{n+1}\biggr).
\end{eqnarray*}
The induction hypothesis gives us $C(b^+)$ and $C(b^-)$; we use that
to eliminate $E(db+)$ and $E(db^-)$ to get
\begin{eqnarray*}
&&b_1b_2E(b_3,\cdots E(b_3,\cdots ,b_{n+1})  +
 E(b_1,\cdots,b_{n-1})b_nb_{n+1}\\
&&-b_1\biggl(b_2E(b_3,\cdots,b_{n+1}) +
(-1)^{n+1}E(b_2,\cdots,b_n)b_{n+1} - C(b^+)\biggr)\\
&&+(-1)^n\biggl(b_1E(b_2,\cdots ,b_n) +
(-1)^{n+1}E(b_1,\cdots,b_{n-1})b_n - C(b^-)\biggr)b_{n+1},
\end{eqnarray*}
which is precisely $C(db)$. \hfill $\Box$
\smallskip

Together, the two propositions give us the following result:
\smallskip

\noindent\textbf{Theorem 5.3} \textit{
The cohomology of the Hochshild complex is  equivalent to the cohomology 
of the restricted complex.}

\section{Geometric cochains of the cone $uv = 0$}  

We intend to discover which, if any, geometric properties of the variety
are reflected in the Hochschild complex.

\bigskip
\noindent{\it 6.1. Points.} Consider a collection of ``geometric $n$-cochains"
$q_1 \otimes \cdots \otimes q_n$, where $q_i$ is a point in $M$, and
the pairing
$$
\langle q_1\otimes \cdots \otimes q_n|a_1\otimes\cdots \otimes
a_n\rangle = \delta_{nm}a_1(q_1)\cdots a_n(q_n),
$$
of geometric cochains with Hochschild chains.  
Here  $a_1,\cdots, a_m$ are the same $m$-chains as before, with
$a_i \in {\mathcal{A}}$, the algebra of coordinate functions on 
$M = \mathbb{C}[u,v]/uv$. 

The  differential is $dq_1 = q_1\otimes q_1$ and
$$
d(q_1\otimes q_2\otimes \cdots) = (q_1\otimes q_1) \otimes q_3\otimes 
\cdots - q_1\otimes (q_2\otimes q_2) \otimes q_3\otimes \cdots + \cdots \,\,;
$$
this ensures the relation of duality,
$$
\langle dq|a\rangle = \langle p|da\rangle.
$$
In this complex of points there is no cohomology. Notice that these
cochains are distributions (delta-functions) valued in $\mathbb{C}$.

The intuitive meaning of this is that the points do not contain any
information about those geometric properties of the  variety that
are related to the existence of deformations. Instead, the tangent vectors
do contain such information. The tangent space is two-dimensional at the
singular point, one-dimensional elsewhere.

\bigskip

\noindent{\it 6.2. Points and tangent vectors.} Let $p_i $ denote  a
tangent vector  at the point  $ q_i$, and consider chains
made up of both $q$'s and $p$'s. The pairing is
$$
\langle q_1\otimes p_2\otimes \cdots|a_1\otimes a_2\otimes
\cdots\rangle = a_1(q_1)(p a_2)(q_2)\cdots.
$$
To preserve duality we define
\begin{eqnarray*}
&&d(p_1\otimes q_2\otimes \cdots) = dp_1\otimes q_2\otimes \cdots -
p_1 \otimes (dq_2)\otimes \cdots + \cdots,\\
&&dq_1 = q_1 \otimes q_1,\quad dp_2 = p_2\otimes q_2 + q_2\otimes
p_2.
\end{eqnarray*}
Let subscript zero refer to the origin, then $q_0 = 0$ and  $p_0\otimes
p_0$ is closed. The 2-cohomology is carried by $p_0\otimes p'_0$.
In particular,
$$
\langle \partial_u\otimes \partial_v|u\otimes v\rangle= 1.
$$

\noindent{\it Conclusion.}  The existence of the antisymmetric part of $H^2$
 on the 2-cone is clearly  a reflection of the fact that the
dimension of tangent space is discontinuous at the singularity, but I
cannot say that the existence of the symmetric part has been clarified in
any deep sense. I hope to improve this type of analysis and to 
generalize it for algebraic manifolds in general.
 
\section{Homology of cones}   

Consider the algebra $\mathbb{C}[x_1,\cdots,x_n]/g$,
where $g = g(x)$ is a second order, homogeneous polynomial,
$$
g = g^{ij}x_ix_j.
$$

The proof of the following is similar to that of Theorem 4.1; what is used 
is the fact that the relations are second order, homogeneous.
\medskip

\noindent\textbf{Theorem 7.1.}  \textit{
Every  chain is cohomologous to a linear chain, and no linear chain is exact, 
so $H_k$ can be identified with the space of closed, linear chains. }
\medskip

\noindent\textbf{Proposition 7.2.} \textit{ 
Let $L_k$ denote the space of alternating, linear $k$-chains, 
then $H_1 = L_1$ and for $k \geq 2, H_k = \oplus_{p\leq k/2}L_{k-2p}$.}
\medskip

\noindent{\it Examples}. For small $k$ the coefficients of representatives
of the homology are
\begin{eqnarray*}
&&k = 2: \alpha^{ij} + \alpha g^{ij},\\
&&k = 3: \alpha^{ijk} + \alpha^ig^{jk} - \alpha^jg^{ik} +
\alpha^kg^{ij},\\
&&k = 4: \alpha^{ijkl} + \alpha^{kl}g^{ij} + \alpha^{il}g^{jk} -
\alpha^{jl}g^{ik} + \alpha^{ij}g^{kl} - \alpha^{kj}g^{li} +
\alpha^{ki}g^{lj}\\
&&\hskip17.5mm  + \alpha(g^{il}g^{jk} - g^{ik}g^{jl} +
g^{ij}g^{kl}), 
\end{eqnarray*}
where the  $\alpha$'s are complex, alternating.

\section{Cohomology of cones} 

Exactly as in the case of the 2-cone, one proves that 
$H^*({\mathrm{Hom}}({\mathcal{A}}^{\otimes n},{\mathcal{A}}))$ 
is the same as for the restricted complex based on linear, closed
chains. For example:
\medskip

Restricted one-forms are closed (never exact) if
$$
dC(x_i\wedge x_j) = 0 = g^{ij}dC(x_i\otimes x_j),
$$
which reduces to $g^{ij}x_iC(x_j) = 0$.
\medskip

Restricted 2-forms are defined for arguments $x_i\wedge x_j$ and
$g^{ij} x_i\otimes x_j$. A calculation shows that a 2-form is closed if
$$
g^{ij}x_iC(x_j\wedge x_k) = 0,
$$
and exact if it is symmetric and valued in ${\mathcal{A}}$.
\medskip

Restricted 3-forms are defined on $x_i\wedge x_j\wedge x_k$ and on
$$
\bar \alpha := (\alpha^ig^{jk} - \alpha^jg^{ik} +
\alpha^kg^{ij})(x_i\otimes x_j \otimes x_k).
$$
They are closed if
$$
g^{ij}x_i C(x_j\wedge  x_k\wedge  x_l) = 0
$$
and 
$$
(g^{il}g^{jk} -g^{jl}g^{ik} +g^{kl}g^{ij})C(x_j\otimes x_k \otimes x_l) = 0,
$$
and exact if  they vanish on $x_i\wedge x_j \wedge x_k$ and, for some
$Q^i \in  {\mathcal{A}}$, $C(\bar \alpha) = \sum_i x_i Q^i.$
\medskip

\noindent\textbf{Acknowledgements}. 
I thank Daniel Sternheimer for useful suggestions,  Murray
Gerstenhaber for information about Harrison cohomology, Didier Arnal for
a useful remark and Georges Pinczon for leading me to cones.

\end{document}